\documentclass[]{aastex631}

\hypersetup{linkcolor=red,citecolor=blue,filecolor=cyan,urlcolor=magenta}

\def\lessim{\mathrel{\hbox{\rlap{\hbox{\lower4pt\hbox{$\sim$}}}\hbox{$<$}}}}
\def\grtsim{\mathrel{\hbox{\rlap{\hbox{\lower4pt\hbox{$\sim$}}}\hbox{$>$}}}}

\usepackage{CJK}

\shorttitle{M31N 2017-01e}
\shortauthors{Shafter et al.}

\graphicspath{{./}{figures/}}

\begin{document}
\begin{CJK*}{UTF8}{gbsn}
\title{Discovery of Two New Eruptions of the Ultrashort Recurrence Time Nova M31N 2017-01e\footnote{Based on observations obtained with the Hobby-Eberly Telescope (HET), which is a joint project of the University of Texas at Austin, the Pennsylvania State University, Ludwig-Maximillians-Universitaet Muenchen, and Georg-August Universitaet Goettingen. The HET is named in honor of its principal benefactors, William P. Hobby and Robert E. Eberly.}}

\correspondingauthor{A. W. Shafter}
\email{ashafter@sdsu.edu}

\author[0000-0002-1276-1486]{Allen W. Shafter}
\affiliation{Astronomy Department, San Diego State University, San Diego, CA 92182, USA}

\author[0000-0002-2770-3481]{Jingyuan Zhao (赵经远)}
\affiliation{Xingming Observatory, Mount Nanshan, Xinjiang, China}

\author[0000-0002-0835-225X]{Kamil Hornoch}
\affiliation{Astronomical Institute of the Czech Academy of Sciences, Fri\v{c}ova 298, CZ-251 65 Ond\v{r}ejov, Czech Republic}

\author[0000-0002-1330-1318]{Hana Ku\v{c}\'akov\'a}
\affiliation{Astronomical Institute, Charles University, Faculty of Mathematics and Physics, V Hole\v{s}ovi\v{c}k\'ach 2, CZ-180 00 Prague, Czech Republic}

\author[0000-0002-8482-8993]{Kenta Taguchi}
\affiliation{Okayama Observatory, Kyoto University, 3037-5 Honjo, Kamogata-cho, Asakuchi, Okayama 719-0232, Japan}

\author[0009-0008-8416-4104]{Jiashuo Zhang (张家硕)}
\affiliation{Xingming Observatory, Mount Nanshan, Xinjiang, China}

\author{Jia You (游嘉)}
\affiliation{Xingming Observatory, Mount Nanshan, Xinjiang, China}

\author[0009-0005-5854-2341]{Binyu Wang (王彬羽)}
\affiliation{Xingming Observatory, Mount Nanshan, Xinjiang, China}

\author{Runwei Xu (许润玮)}
\affiliation{Xingming Observatory, Mount Nanshan, Xinjiang, China}

\author[0009-0000-5910-0004]{Weiye Wang (王炜晔)}
\affiliation{Xingming Observatory, Mount Nanshan, Xinjiang, China}

\author[0009-0004-3336-7842]{Yuqing Ren (任育庆)}
\affiliation{Xingming Observatory, Mount Nanshan, Xinjiang, China}

\author[0009-0009-0237-5133]{Lanhe Ding (丁岚贺)}
\affiliation{Xingming Observatory, Mount Nanshan, Xinjiang, China}

\author[0009-0003-8247-791X]{Xiaochang Yan (严小畅)}
\affiliation{Xingming Observatory, Mount Nanshan, Xinjiang, China}

\author[0000-0003-3005-2189]{Mi Zhang (张宓)}
\affiliation{Xingming Observatory, Mount Nanshan, Xinjiang, China}

\author[0000-0003-2588-1265]{Wei-Hao Wang (王为豪)}
\affiliation{Institute of Astronomy and Astrophysics, Academia Sinica, Taipei 10617, Taiwan}

\author[0000-0003-1377-7145]{Howard E. Bond}
\affiliation{Department of Astronomy \& Astrophysics, Penn State University, University Park, PA 16802, USA}
\affiliation{Space Telescope Science Institute, 3700 San Martin Dr., Baltimore, MD 21218, USA}

\author[0000-0002-3742-8460]{Robert Williams}
\affiliation{Space Telescope Science Institute, 3700 San Martin Dr., Baltimore, MD  21218, USA}
\affiliation{Department of Astronomy \& Astrophysics, Univ. of California Santa Cruz, 1156 High Street,
Santa Cruz, CA 95064}

\author[0000-0003-2307-0629]{Gregory R. Zeimann}
\affiliation{Department of Astronomy, University of Texas, Austin, TX, 78712, USA}

\begin{abstract}

We report the recent discovery of two new eruptions of the recurrent nova M31N 2017-01e in the Andromeda galaxy. The latest eruption, M31N 2024-08c, reached $R=17.8$ on 2024 August 06.85 UT, $\sim2$ months earlier than predicted. In addition to this recent eruption, a search of archival PTF data has revealed a previously unreported eruption on 2014 June 18.46 UT that reached a peak brightness of $R\sim17.9$ approximately a day later.
The addition of these two eruption timings
has allowed us to update the mean recurrence time of the nova. We find $\langle T_\mathrm{rec} \rangle = 924.0\pm7.0$~days ($2.53\pm0.02$~yr), which is slightly shorter than our previous determination. Thus, M31N 2017-01e remains the nova with the second shortest recurrence time known, with only M31N 2008-12a being shorter. We also present a low-resolution spectrum of the likely quiescent counterpart of the nova, a $\sim20.5$ mag evolved B star displaying an $\sim14.3$~d photometric modulation.

\end{abstract}


\section{Introduction}

\begin{figure*}
\includegraphics[angle=0,scale=0.2]{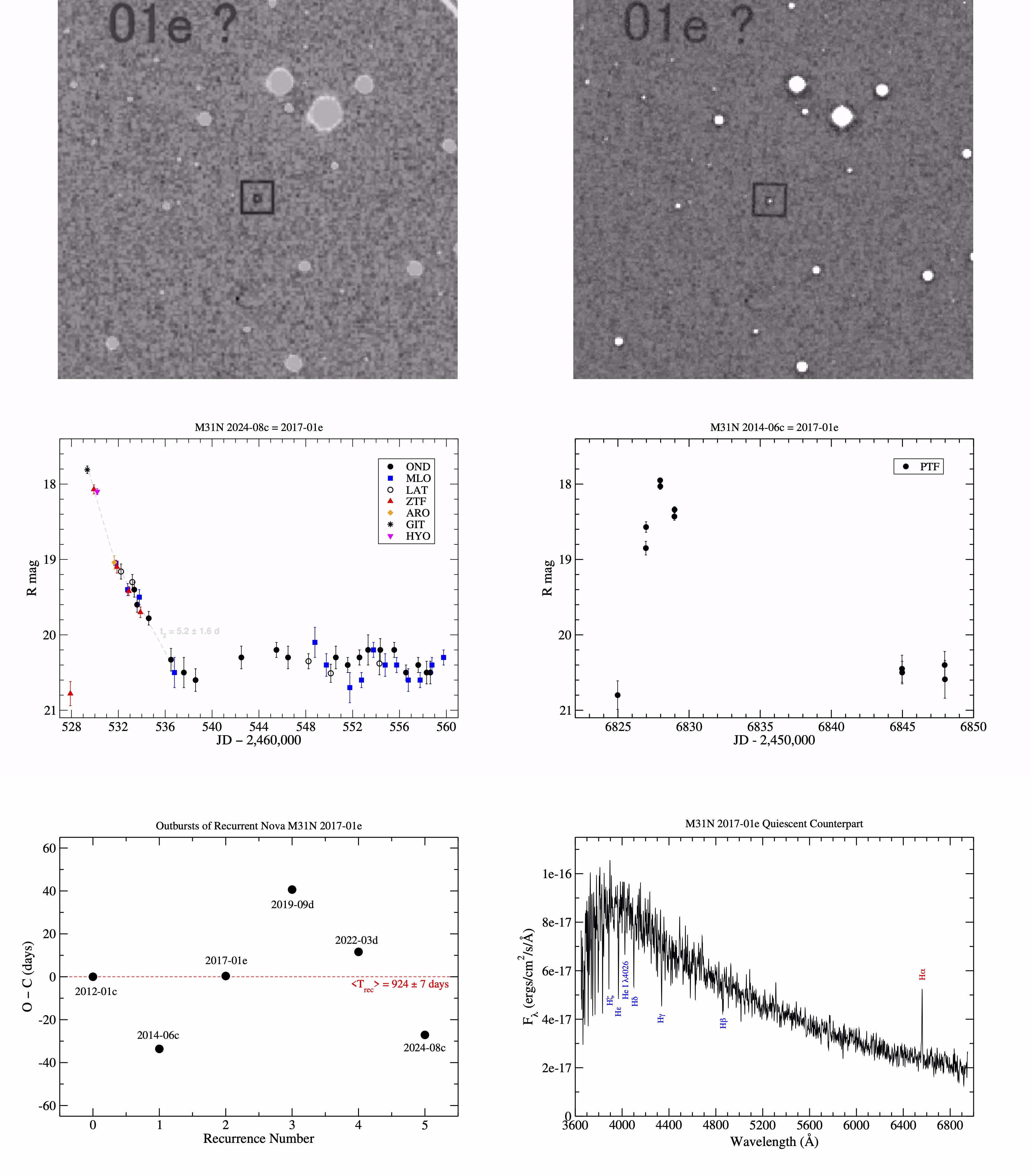}
\caption{
{\it Top Panel:} The positions of M31N 2024-08c and 2014-06c (white) are compared with 2017-01e (black) in the left and right panels, respectively. The positions of the novae are coincident to an accuracy of less than $1''$.
{\it Middle panel:} The lightcurves for M31N 2024-08c and 2014-06c are shown in the left and right panels, respectively (see {\it Facilities\/} section for symbol key and the Appendix for photometry). The 2024-08c lightcurve shows that $t_2\sim$5~d, which is consistent with other observed eruptions. The GIT point is from \citet{2024ATel16759....1B}.
{\it Bottom left panel:} The $O-C$ plot showing the times of the 6 observed eruptions of M31N 2017-01e.
{\it Bottom right panel:} Our HET spectrum of the putative quiescent optical counterpart, appearing to show a luminous B-type star.
%(Photometry of M31N 2014-06c and 2024-08c is available in the DBF files.)
}
\label{f1}
\end{figure*}

Recently, we reported observations of an unusual recurrent nova system in the Andromeda galaxy, M31N 2017-01e, establishing that
this nova has a recurrence time of $\sim 2.5$~yr, which is the second shortest recurrence time known for any nova \citep{2022RNAAS...6..241S}.
We also established that
the quiescent counterpart of the nova is likely associated with a blue ($B-V\sim0.045$), $V\sim20.5$ mag variable star that displays a 14.3~d photometric modulation. Given that the nova typically reaches $R\sim17.8$ ($M_R\sim-6.7$) at maximum light, this implies an eruption amplitude of only $\sim3$~mag.
Furthermore, at the distance of M31, we find $M_V\sim-4$ at quiescence, which, given its blue color, would be unprecedented for any known recurrent nova system.
Here, we
report additional observations of this remarkable nova, including the discovery of its most recent eruption\footnote{\tt see http://www.cbat.eps.harvard.edu/unconf/followups/J00441071+4154221.html} (M31N 2024-08c), as well as the discovery in archival Palomar Transient Factory (PTF) data of a previously unidentified eruption in 2014 June (M31N 2014-06c). We also present for the first time a low-resolution spectrum of the putative quiescent optical counterpart.

\section{Discovery of Two Eruptions: One New, One Old}

The recent transient near the position of M31N 2017-01e was detected on 2024 August 06.85 UT. A careful overlay of the position of this new transient with that of M31N 2017-01e (see Fig.~\ref{f1}) establishes that the two events arise from the same progenitor system. A complete light curve of the eruption, also shown in Figure~\ref{f1}, establishes that the nova reached $R\sim17.8$ before fading rapidly ($t_2 = 5.2\pm1.6$~d)\footnote{The $t_2$ time is the time in days for a nova to fade by 2 magnitudes.} back to quiescence near $R=20.5$.

Spurred by the recent eruption, a search of the PTF archive by one of us (JZ) has also revealed a previously unrecognized $R=17.9$ mag transient peaking on 2014 June 19.47 UT near the position of M31N 2017-01e. A partial light curve of this previously unknown transient is also presented in Figure~\ref{f1}, along with 
a careful alignment of the PTF image at maximum light with that of M31N 2017-01e. As with the latest eruption, the two events are spatially
coincident to better than an arcsec establishing once again that the two events arise from the same progenitor.

\subsection{Updated Mean Recurrence Time}

With the addition of the times of the 2nd and 6th observed eruptions (MJD 56826.46 and MJD 60528.85 for the 2014 June and 2024 August eruptions, respectively) to the four timings given in \citet{2022RNAAS...6..241S}\footnote{Based on PTF data, we have revised the time of the 2012 eruption to MJD 55936.11, about a day earlier than reported previously.}, we find 
$\langle T_\mathrm{rec}\rangle=924.0\pm7.0$~d ($2.53\pm0.02$ yr), with
the full ephemeris given by: MJD$_\mathrm{erupt} = (55936.1\pm1.0) + (924.0\pm7.0)$~E. The observed and expected (computed) times of the 6 eruptions are
compared in the bottom left panel of Figure~\ref{f1}.
Based on the regularity of the eruptions
and our updated ephemeris, we expect the next eruption of M31N 2017-01e to occur in 2027 March, give or take about a month. Given that M31 will be in conjunction with the sun then, it is quite possible that this eruption will unfortunately be missed.

\section{A Spectrum of the Putative Quiescent Counterpart}

\citet{2022RNAAS...6..241S} discussed possible models for the nova, pointing out that
the quiescent counterpart appears to be
a $\sim20.5$ mag variable source with a strong ($\sim0.5$ mag) photometric modulation \citep{2006A&A...459..321V}.
In an attempt to further characterize the quiescent counterpart, on 2023 January 25 we obtained a spectrum with the low-resolution spectrograph (LRS2-B)\footnote{For reductions see: \url{https://github.com/grzeimann/Panacea};
\url{https://github.com/grzeimann/LRS2Multi}} on the Hobby-Eberly Telescope (HET) \citep{1998SPIE.3352...34R, 2021AJ....162..298H}.
The spectrum, shown in the lower right panel of Figure~\ref{f1}, reveals narrow Balmer and He~I ($\lambda=4026$\AA) absorption features coupled with similarly narrow H$\alpha$ emission (FWHM $\lessim150$ km~s$^{-1}$), superimposed on a blue continuum. The spectrum is consistent with being dominated by an evolved B star; however it is not obvious how such a star would be associated
with a nova progenitor. In particular, it would have a mass considerably greater than that of a Chandrasekhar-mass WD making mass transfer onto the white dwarf via Roche lobe overflow likely to be dynamically unstable \citep[]{2023A&A...669A..45T}. On the other hand, mass transfer via a stellar wind would seem unlikely to provide the accretion rate necessary ($\dot M \grtsim 10^{-7}$~M$_\odot$~yr$^{-1}$) to trigger nova eruptions with a 2.5~yr recurrence time \citep[e.g., see][]{2014ApJ...793..136K}. A complete model for the M31N 2017-01e system will require a more thorough investigation of the quiescent counterpart, which is currently underway.

\begin{acknowledgments}
JZ thanks the Alnitak Remote Observatories ARTA program for observing time.
KH was supported by the project RVO:67985815.
We are grateful to D. Schneider for DD time on the HET.
\end{acknowledgments}

\facilities{Alnitak Remote Observatories (ARO); Growth India Telescope (GIT); Hobby-Eberly Telescope (HET); Huanyu Observatory (HYO); Lulin-ASIAA Telescope for Transients and Education (LAT); Mount Laguna Observatory (MLO); Ond\v{r}ejov Observatory (OND); Palomar Transient Factory (PTF); Zwicky Transient Facility (ZTF).}

\newpage

\bibliography{M31N2017-01e}{}
\bibliographystyle{aasjournal}

\appendix

\startlongtable
\begin{deluxetable}{clcr}
%\tabletypesize{\scriptsize}
\tablenum{A1}
\tablecolumns{4}
\tablecaption{Nova Photometry}
\tablehead{\colhead{$\mathrm{JD}~(+ 2,400,000$)} & \colhead{Filter\tablenotemark{a}} & \colhead{Mag} & \colhead{Telescope\tablenotemark{b}}
}
\startdata
\cutinhead{M31N 2014-06c}
56824.9788&R &$20.80\pm0.19$&PTF\cr
56826.9642&R &$18.85\pm0.09$&PTF\cr
56826.9776&R &$18.57\pm0.07$&PTF\cr
56827.9667&R &$17.95\pm0.03$&PTF\cr
56827.9791&R &$18.03\pm0.04$&PTF\cr
56828.9682&R &$18.43\pm0.05$&PTF\cr
56828.9808&R &$18.34\pm0.04$&PTF\cr
56844.9679&R &$20.45\pm0.18$&PTF\cr
56844.9804&R &$20.50\pm0.15$&PTF\cr
56847.9678&R &$20.40\pm0.18$&PTF\cr
56847.9829&R &$20.59\pm0.25$&PTF\cr
\cutinhead{M31N 2024-08c}
60527.906&Rr&$20.78\pm0.16$& ZTF\cr
60529.350&Rr&$17.81\pm0.05$& GIT\cr
60529.913&Rr&$18.07\pm0.06$& ZTF\cr
60530.196&R &$18.10\pm0.04$& HYO\cr
60531.662&R &$19.04\pm0.09$& ARO\cr
60531.796&R &$19.05\pm0.04$& MLO\cr
60531.882&Rr&$19.10\pm0.08$& ZTF\cr
60532.230&R &$19.16\pm0.03$& LAT\cr
60532.795&R &$19.40\pm0.08$& MLO\cr
60532.889&Rr&$19.42\pm0.06$& ZTF\cr
60533.194&R &$19.30\pm0.10$& LAT\cr
60533.360&R &$19.40\pm0.10$& OND\cr
60533.604&R &$19.60\pm0.10$& OND\cr
60533.792&R &$19.50\pm0.10$& MLO\cr
60533.885&Rr&$19.70\pm0.07$& ZTF\cr
60534.595&R &$19.78\pm0.09$& OND\cr
60536.497&R &$20.33\pm0.15$& OND\cr
60536.791&R &$20.50\pm0.20$& MLO\cr
60537.596&R &$20.50\pm0.20$& OND\cr
60538.589&R &$20.60\pm0.15$& OND\cr
60542.500&R &$20.30\pm0.15$& OND\cr
60545.496&R &$20.20\pm0.10$& OND\cr
60546.486&R &$20.30\pm0.15$& OND\cr
60548.233&R &$20.35\pm0.10$& LAT\cr
60548.781&R &$20.10\pm0.20$& MLO\cr
60549.757&R &$20.40\pm0.15$& MLO\cr
60550.134&R &$20.51\pm0.12$& LAT\cr
60550.567&R &$20.30\pm0.15$& OND\cr
60551.580&R &$20.40\pm0.10$& OND\cr
60551.758&R &$20.70\pm0.20$& MLO\cr
60552.583&R &$20.30\pm0.10$& OND\cr
60552.755&R &$20.60\pm0.10$& MLO\cr
60553.321&R &$20.20\pm0.20$& OND\cr
60553.778&R &$20.20\pm0.10$& MLO\cr
60554.288&R &$20.38\pm0.15$& LAT\cr
60554.360&R &$20.20\pm0.15$& OND\cr
60554.776&R &$20.40\pm0.15$& MLO\cr
60555.549&R &$20.20\pm0.10$& OND\cr
60555.753&R &$20.40\pm0.10$& MLO\cr
60556.554&R &$20.50\pm0.10$& OND\cr
60556.750&R &$20.60\pm0.15$& MLO\cr
60557.595&R &$20.40\pm0.10$& OND\cr
60557.759&R &$20.60\pm0.10$& MLO\cr
60558.315&R &$20.50\pm0.15$& OND\cr
60558.634&R &$20.50\pm0.15$& OND\cr
60558.770&R &$20.40\pm0.10$& MLO\cr
60559.761&R &$20.30\pm0.10$& MLO\cr
\enddata
\tablenotetext{a}{ R -- Cousins $R$; Rr -- Sloan $r$ converted to Cousins $R$}
\tablenotetext{b}{
          ARO -- Alnitak Remote Observatories;
          GIT -- Growth India Telescope;
          HYO -- Huanyu Observatory;
          LAT -- Lulin-ASIAA Telescope for Transients and Education;
          MLO -- Mount Laguna Observatory;
          OND -- Ond\v{r}ejov Observatory;
          PTF -- Palomar Transient Factory;
          ZTF -- Zwicky Transient Facility.}
\end{deluxetable}

\end{CJK*}
\end{document}